\documentclass{aa}  
\usepackage{graphicx}
\usepackage{txfonts,color}
\usepackage{natbib}
\bibpunct{(}{)}{;}{a}{}{,} 
\usepackage{CJKutf8}
\usepackage{multirow}

\usepackage[colorlinks,     citecolor = blue]{hyperref}

\newcommand{\HII}{\ion{H}{ii}}
\newcommand{\HI}{\ion{H}{i}} 

\newcommand{\lzifu} {{\scshape lzifu}}
\newcommand{\ppxf} {{\scshape ppxf}}

\newcommand{\izi} {{\scshape izi}}
\newcommand{\mappingsiv} {{\scshape mappings~iv}}
\newcommand{\mappingsv} {{\scshape mappings~v}}
\newcommand{\hiiphot} {{\scshape hiiphot}}

\begin{document} 
\begin{CJK*}{UTF8}{bkai}

   \title{Azimuthal variations of gas-phase oxygen abundance in NGC\,2997}
  \subtitle{}
\titlerunning{Azimuthal variations of gas-phase oxygen abundance in NGC\,2997}
\authorrunning{Ho et al.}

   \author{I-Ting Ho (何宜庭)
          \inst{1}
          \and
          Sharon E. Meidt
          \inst{1}
          \and 
          Rolf-Peter Kudritzki
          \inst{2,3}
          \and
          Brent A. Groves
          \inst{4}
          \and 
          Mark Seibert
          \inst{5}
          \and
          Barry F. Madore
          \inst{5}
          \and
          Eva Schinnerer
          \inst{1}
          \and
          Jeffrey A. Rich
          \inst{5}
          \and
          Chiaki Kobayashi
          \inst{6}
          \and
          Lisa J. Kewley
          \inst{5}
          }

\institute{Max Planck Institute for Astronomy, K\"{o}nigstuhl 17, 69117 Heidelberg, Germany
\and
Institute for Astronomy, University of Hawaii, 2680 Woodlawn Drive, Honolulu, HI 96822, USA
\and
University Observatory Munich, Scheinerstr. 1, D-81679 Munich, Germany
\and
Research School of Astronomy and Astrophysics, Australian National University, Canberra 2600, ACT, Australia
\and
Carnegie Observatories, 813 Santa Barbara Street, Pasadena, CA 91101, USA
\and
Centre for Astrophysics Research, School of Physics, Astronomy and Mathematics, University of Hertfordshire, College Lane, Hatfield AL10 9AB, UK
}

   \date{}

 
  \abstract
  {The azimuthal variation of the \HII\ region oxygen abundance in spiral galaxies is a key observable for understanding how quickly oxygen produced by massive stars can be dispersed within the surrounding interstellar medium. Observational constraints on the prevalence and magnitude of such azimuthal variations remain rare in the literature. Here, we report the discovery of pronounced azimuthal variations of \HII\ region oxygen abundance in NGC\,2997, a spiral galaxy at approximately 11.3~Mpc. Using 3D spectroscopic data from the TYPHOON Program, we study the \HII\ region oxygen abundance at a physical resolution of 125~pc. Individual \HII\ regions or complexes are identified in the 3D optical data and their strong emission line fluxes measured to constrain their oxygen abundances. We find 0.06~dex azimuthal variations in the oxygen abundance on top of a radial abundance gradient that is comparable to those seen in other star-forming disks. At a given radial distance, the oxygen abundances are highest in the spiral arms and lower in the inter-arm regions, similar to what has been reported in NGC\,1365 using similar observations. We discuss whether the azimuthal variations could be recovered when the galaxy is observed at worse physical resolutions and lower signal-to-noise ratios. }


   \maketitle
%

\section{Introduction}

The oxygen abundance in the interstellar medium (ISM) in an important astrophysical quantity for the understanding of galaxy formation and evolution. The oxygen is primarily synthesized inside massive stars before being released to the ISM when massive stars explode as supernovae. Thus, the ISM oxygen abundance contains a fossil record of the mass assembly history of a galaxy. The oxygen abundance around \HII\ regions can be easily measured using bright emission lines observable in the optical window. These attractive features have made using \HII\ region oxygen abundances to study galaxy evolution an active area of observational and theoretical research \citep[e.g.][]{Tremonti:2004fk,Kobayashi:2007fk,Dave:2011qy,Lilly:2013qf,Zahid:2014uq}.

The spatial distribution of the oxygen abundance in galaxies provides a critical constraint in understanding the build-up of galactic disks. Since the early work by \citet{Searle:1971kx}, subsequent observations have established the prevalence of oxygen abundance gradients in star-forming disks \citep[e.g.][]{Vila-Costas:1992fj,Zaritsky:1994lr,Moustakas:2010fk}. The negative slopes of the gradients are believed to reflect the inside-out formation history of galaxies \citep[e.g.][]{Chiappini:1997fk,Chiappini:2001fk,Fu:2009vn}. Recent progress in integral field spectroscopy (IFS), or 3D spectroscopy, further solidifies the existence of a ``common'' abundance gradient in all local star-forming disks \citep[][see also \citealt{Sanchez-Menguiano:2018aa}]{Sanchez:2014fk,Ho:2015hl}. The statistical power from these new datasets also allows high-precision quantifications of the slope, and investigating higher-order correlations with other physical parameters, for example stellar mass, bar, interaction stage \citep[e.g.][]{Belfiore:2017aa,Sanchez-Menguiano:2016vl}. 
\begin{table}
 \caption{NGC\,2997}
 \label{tbl1}
\centering
\begin{tabular}{lcl}
  \hline
  Property & Value & Note \\
  \hline
  Distance$^a$ & 11.3~Mpc & $1\rm \arcsec = 55.8~pc;\ 1\arcmin = 3.3~kpc.$ \\
  Inclination$^b$ & $40^\circ$ &  --\\
  P.A.$^b$ & $110^\circ$ & -- \\
  $\rm R_{25}$$^c$ & 4.46\arcmin & B-band$^c$ \\
  \hline
\multicolumn{3}{p{\columnwidth}}{
$^a$~Extragalactic Distance Database. \href{http://edd.ifa.hawaii.edu/ }{http://edd.ifa.hawaii.edu/}.
$^b$~\citet{Milliard:1981aa}
$^c$~RC3 \citep{de-Vaucouleurs:1991sf}}
 \end{tabular}
\end{table}

While the radial gradients in star-forming disks are routinely measured, azimuthal variations of abundances have been less well studied. The azimuthal variations reflect how quickly the oxygen produced by star-forming activities is mixed with the surrounding medium and subsequently distributed around the galactic disk when these recently processed elements move within the gravitational potential of the system \citep{Roy:1995ys}. Evidence suggesting that Milky Way ISM abundances depend on the azimuthal angle have been found using \HII\ regions \citep{Balser:2011uo} and red supergiant clusters \citep[][see also the studies with Cepheids by \citealt{Pedicelli:2009qq,Genovali2014}]{Davies:2009aa}. However, beyond the Milky Way, most early spectroscopic studies of \HII\ regions in nearby, non-interacting galaxies have revealed little evidence for azimuthal variations \citep[e.g.][]{Martin:1996nr,Cedres:2002vl,Bresolin:2011rt}, although tentative evidence has been reported in some studies \citep{Kennicutt:1996yf,Rosolowsky:2008qf,Li:2013nx}. Recent 3D spectroscopy studies covering the entire star-forming disks have started to reveal the presence of azimuthal variations of oxygen abundance \citep[][see also \citealt{Croxall:2016tg}]{Sanchez-Menguiano:2016zr,Vogt:2017aa,Ho:2017aa,Sakhibov:2018aa}, although not all studies have found measurable variations \citep[][see also \citealt{Zinchenko:2016aa} who studied azimuthal asymmetry]{Kreckel:2016aa}. The magnitude of these variations also differs from study to study (e.g. peak-to-peak variations of 0.4~dex in \citealt{Ho:2017aa}, 0.1~dex in \citealt{Sanchez-Menguiano:2016zr}, 0.1 to 0.3~dex in \citealt{Vogt:2017aa}, and 0.01 to 0.06~dex in \citealt{Sakhibov:2018aa}).

\begin{figure*}
\centering
\includegraphics[width=\textwidth]{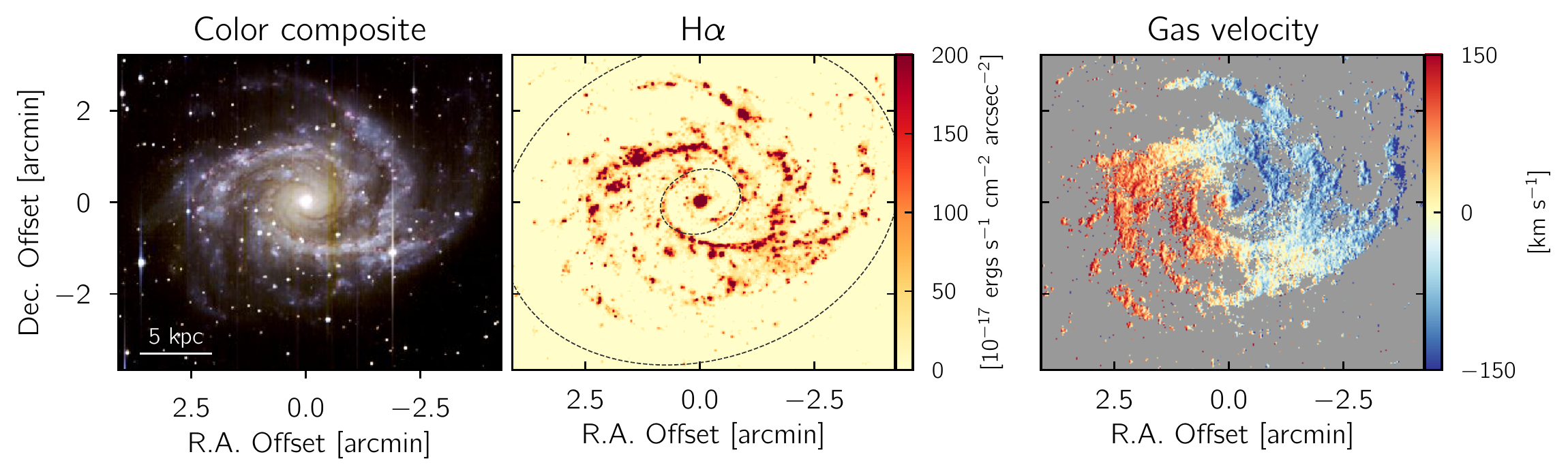}
\caption{
{\it Left}: {\it BVR} color composite image of NGC\,2997 reconstructed from the data cube. 
{\it Middle}: H$\alpha$ map derived from fitting the emission line. The two dashed ellipses correspond to $0.2\times$ and $1\times\rm R_{25}$ in the frame of the galactic disk. This work focuses on \HII\ regions falling in between the two ellipses.  
{\it Right}: Ionized gas velocity field measured from fitting emission lines. 
North is to the top and east to the left. }\label{fig1} 
\end{figure*}

\begin{figure}
\centering
\includegraphics[width=\columnwidth]{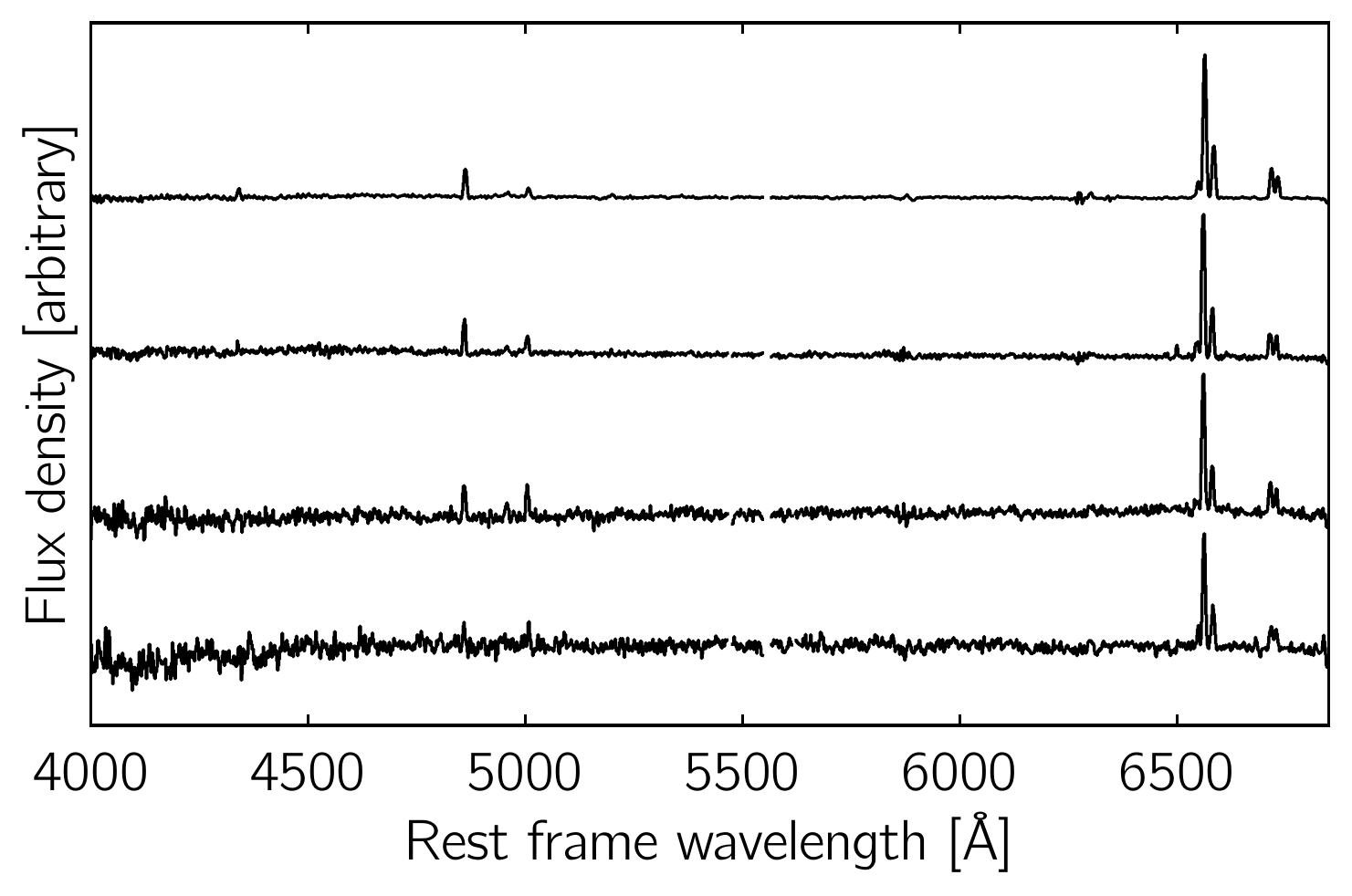}
\caption{Representative \HII\ region spectra from the first to the last S/N(H$
\alpha$) quartiles. The four scaled spectra have S/N(H$
\alpha$) of 237, 119, 68, and 35 (from top to bottom).}\label{fig2}
\end{figure}

Emerging evidence suggest that spiral arms may play a critical role in driving the observed azimuthal variations because many variations discovered correlate spatially with the spiral structure. \citet{Sanchez-Menguiano:2016zr} attribute the spatial correlation between azimuthal variations and spiral arms in NGC\,6754 to radial flows induced by spiral arms (see also \citealt{Sanchez:2015aa}). Such radial flows are commonly seen in numerical simulations \citep{Grand:2016hb}. \citet{Ho:2017aa} argue that the variations in NGC\,1365 are due to the shortened mixing time when spiral density waves pass through, which increases the cloud-cloud collision rate, star formation rate and subsequently the supernova production rate.

\begin{figure}
\centering
\includegraphics[width=0.95\columnwidth]{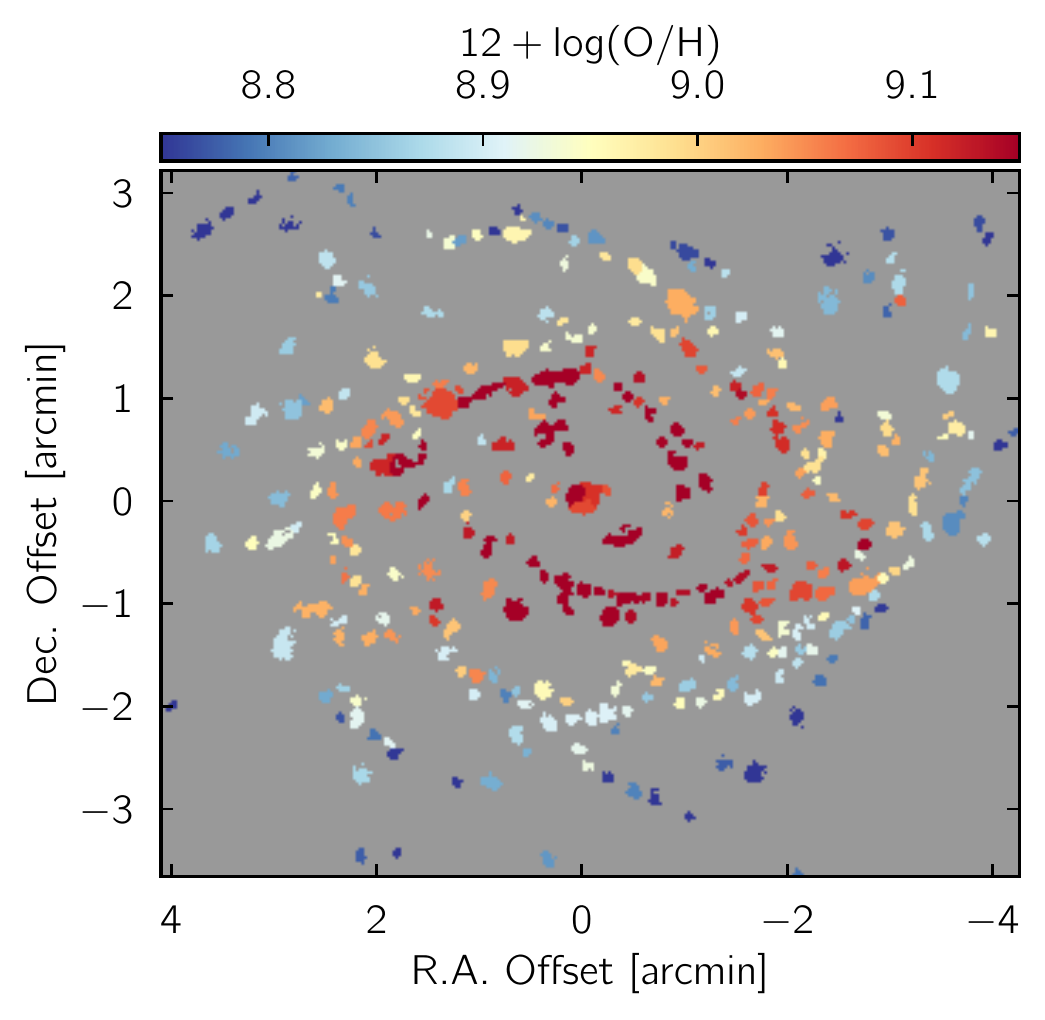}
\caption{Gas-phase oxygen abundance map. Each region on the maps corresponds to one \HII\ region identified by \hiiphot\ on the H$\alpha$ map and has line ratios consistent with photoionization. We note that the abundance scale here is from the D13 model.  The mean oxygen abundance derived from the $T_e$ method is approximately 0.7 solar, as reported by \citet{Bresolin:2005uq} who detected auroral lines in 7 \HII\ regions on the disk.}\label{fig3}
\end{figure}

While the physical mechanisms driving the variations remain unclear, progress has been made to understand the prevalence of azimuthal variations. An important step forward was taken by \citet{Sanchez-Menguiano:2017aa} who analysed arm and inter-arm metallicity gradients in a statistical sample of 63 galaxies from the Calar Alto Legacy Integral Field Area Survey survey \citep[CALIFA;][]{Sanchez:2012fj}. They suggested that galaxy morphology could be an important factor for the presence of azimuthal variations. Flocculent galaxies present different arm and inter-arm abundance gradients while grand design systems do not. However, the typical spatial resolution of 1~kpc reached by the CALIFA survey is still larger than the typical size of an \HII\ region (tens to hundreds parsec). The \HII\ region oxygen abundances derived can thus be affected by the diffuse gas that contaminates the emission line measurements \citep[][Poetrodjojo et al. (in preparation)]{Yuan:2013gf,Mast:2014lr}. This could be particularly important for the inter-arm \HII\ regions. 

The logical next step for investigating the nature of the azimuthal variations is to conduct case studies of nearby galaxies at distances where individual, inter-arm \HII\ regions can be resolved. Here, we report the discovery of pronounced azimuthal variations of oxygen abundance in NGC\,2997
determined from 3D spectroscopic data from the TYPHOON Program (Seibert et al. in preparation), an on-going IFS survey of the largest angular-sized galaxies in the southern hemisphere. The high physical resolution of about 100 to 150 parsec enables the identification of individual \HII\ regions or complexes, and then the accurate derivation of their oxygen abundances. NGC\,2997 is the second galaxy from the TYPHOON Program which shows clear azimuthal variations, following the first galaxy NGC\,1365 presented in \citet{Ho:2017aa}.

\section{Observations and data reduction}
NGC\,2997 was observed using the 2.5~m du Pont telescope at the Las Campanas Observatory in Chile. The observations are a part of the long-term TYPHOON Program that is obtaining optical data cubes for the largest angular-sized galaxies in the southern hemisphere. For more details about the program and its first science results, the reader is directed to Seibert et al. (in preparation) and \citet{Ho:2017aa}. Here, we only provide a short summary.

\begin{figure*}
\centering
\includegraphics[width=0.85\textwidth]{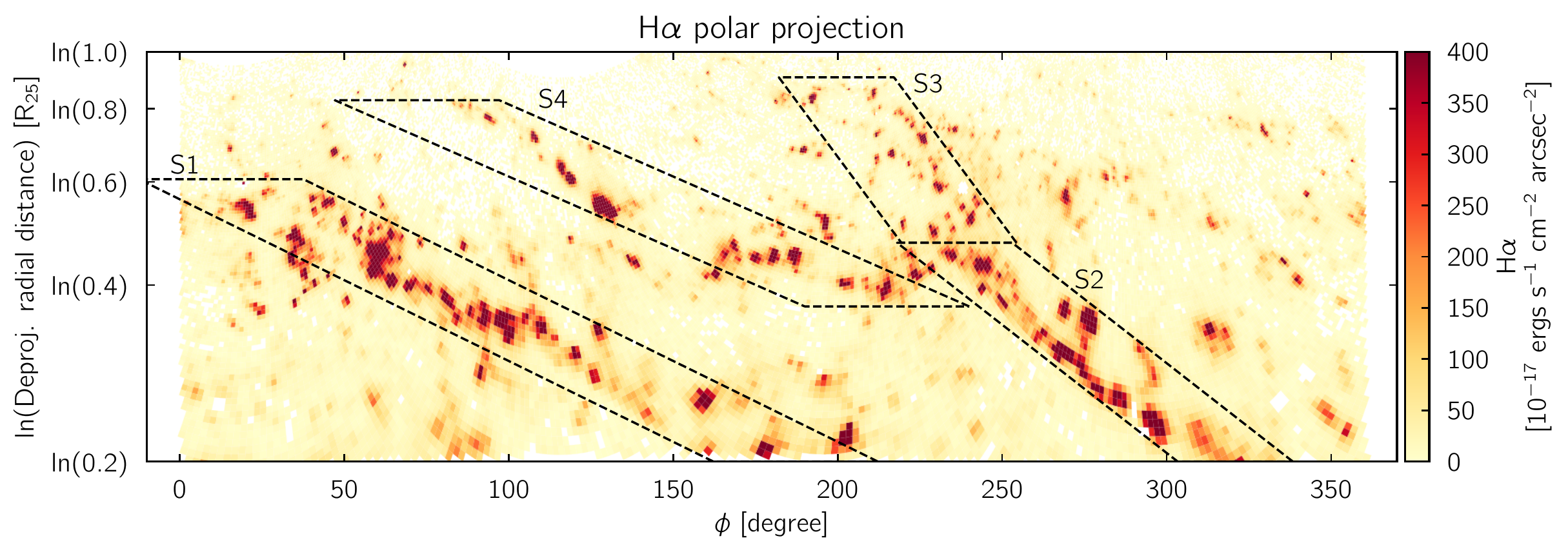}
\caption{H$\alpha$ map in polar projection, assuming a circular thin disk and the geometric parameters in Table~\ref{tbl1}. Linear structures in this $\ln(r)-\phi$ parameter space correspond to logarithmic spirals. Four such structures are identified by eye and marked as S1, S2, S3, and S4. The resulting spiral arm mask in the plane of the sky is shown in Figure~\ref{fig5}. Details are described in Section~3.2.}\label{fig4} 
\end{figure*}

\begin{figure}
\centering
\includegraphics[width=\columnwidth]{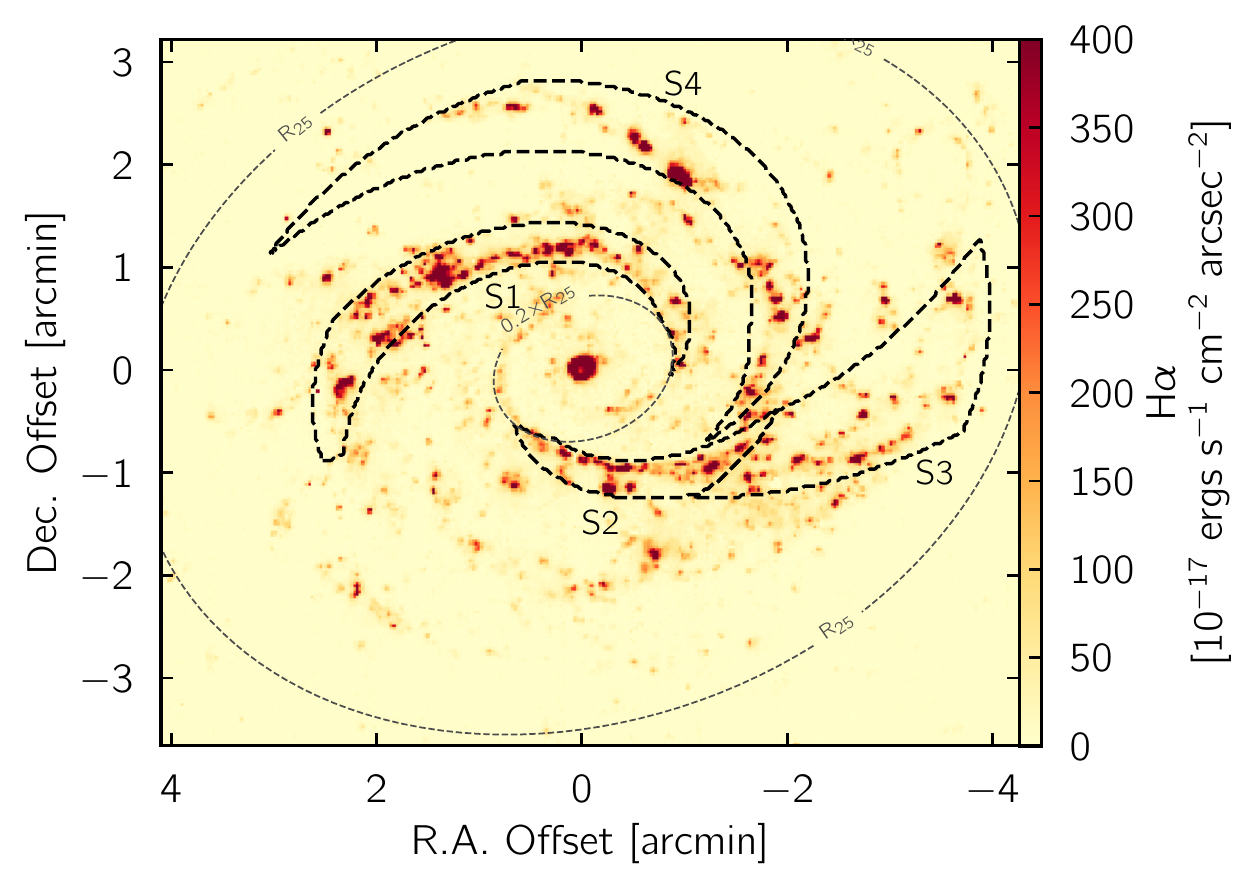}
\caption{Spiral arm mask as determined from Figure~\ref{fig4} (polar projection). Details are described in Section~3.2.}\label{fig5} 
\end{figure}

The TYPHOON Program adopts the Progressive Integral Step Method (i.e,~``step-and-stare'' or ``stepped-slit'' techniques) to build up 3D data cube using a long slit. The focal plane element is a custom-built long-slit of 18\arcmin\ long and 1.\arcsec65 wide. The back end imaging spectrograph, the Wide Field reimaging CCD (WFCCD), was configured to a resolution of approximately 3.5 \AA\ ($\sigma$), corresponding to a resolving power $R\ (\equiv\lambda/\delta\lambda)\approx 850$ at 7000\AA.

During the NGC\,2997 observations, the long-slit was positioned in the north-south direction and progressively scanned across the galaxy. Each slit position was integrated for 600 seconds, then the slit was moved by one slit-width for the next integration. In total, 304 integrations or equivalently 50.7 hours were conducted during 4 observing runs spanning a total of 20 nights in January 2013, April 2013, January 2016, and February 2016. An area of $8.\arcmin 4\times 18\arcmin$ was covered by the observations, encompassing the entire optical disk of NGC\,2997 (R$_{25}$ of 4.46\arcmin). The integration time results in 6000\AA\ continuum, per-channel signal-to-noise ratios of approximately 50 at the galaxy center and 3 at about 0.5 R$_{25}$.

Standard long-slit data reduction techniques were used to calibrate the data. Sky subtraction is performed on each 2D-spectral image (step) independently because each step is non-contemporaneous and sky conditions change. Due to the wide field of view, each 2D-spectral image is rectified to remove non-linear wavelength and spatial distortions. Once the 2D-spectral images are corrected, sky values are taken as the mode in each column of the dispersed 2D image using a bin size of one half the size of the read noise. If the result is not single valued then the bin size is increased by 1\% until it becomes single valued. This method is robust as long as the scanned object does not dominate the 18 arc-minutes of the slit with a uniform flux level. The reduced 2D spectra were then assembled into a 3D data cube. The astrometry accuracy of the final data cube is approximately 4\arcsec, as estimated from field stars. The final data cube covers a wavelength range of 3650 to 8150\AA. The spectral and spatial samplings are 1.5\AA\ and 1.\arcsec65, respectively. The point spread function is approximately 2 to 2.\arcsec5 (FWHM; 110 to 140~pc at the assumed distance; Table~\ref{tbl1}), as determined from fitting field stars with Gaussians. The left panel of Figure~\ref{fig1} presents a BVR color-composite image reconstructed from the 3D data cube.

\section{Data analysis}

\subsection{Derive \HII\ region oxygen abundance from data cube}
To derive oxygen abundances using \HII\ regions and the optical data cube, we follow the same procedures as \citet{Ho:2017aa}. Below, we provide a brief summary. Details about the analysis techniques are described in \citet{Ho:2017aa}. 

We first construct an H$\alpha$ map from the data cube using the emission line fitting tool \lzifu\ \citep{ho:2016rc,ho:2016aa}. The stellar continuum is removed prior to fitting the emission lines using the \ppxf\ code \citep{Cappellari:2004uq,Cappellari:2017aa} and the \textsc{miuscat} simple stellar population models \citep{Vazdekis:2012yq}. The resulting 2D maps from the emission line analysis are shown in Figure~\ref{fig1}. From the H$\alpha$ map, \HII\ region candidates are identified using an adapted version of the \hiiphot\ code \citep{Thilker:2000rf}. About 500 \HII\ region candidates are identified based on the H$\alpha$ surface brightness distribution . We then integrate a single spectrum for each \HII\ region candidate from the data cube using the \HII\ region mask. Emission lines of each \HII\ region candidate are then measured using \lzifu. The emission lines included in the \lzifu\ analysis are: H$\beta$, [\ion{O}{iii}]$\lambda\lambda 4959,5007$, [\ion{N}{ii}]$\lambda\lambda 6548,6583$, H$\alpha$, and  [\ion{S}{ii}]$\lambda\lambda 6716,6731$. Hereafter, we adopt the following convention for line notation: [\ion{O}{iii}] $\equiv$ [\ion{O}{iii}]$\lambda$5007, [\ion{N}{ii}] $\equiv$ [\ion{N}{ii}]$\lambda$6583, and  [\ion{S}{ii}] $\equiv$ [\ion{S}{ii}]$\lambda$6716 +  [\ion{S}{ii}]$\lambda$6731.

Before deriving the oxygen abundance from emission line ratios, \HII\ region candidates presenting low S/N or line ratios inconsistent with photoionization by OB stars are rejected. We adopt a S/N cut of 3 on H$\beta$, [\ion{O}{iii}], H$\alpha$, [\ion{N}{ii}], and [\ion{S}{ii}]$\lambda\lambda 6716,6731$, and exclude regions inconsistent with \HII\ region based on their [\ion{N}{ii}]/H$\alpha$ and [\ion{O}{iii}]/H$\beta$ line ratios. 
The empirical separation line by \citet{Kauffmann:2003vn} is adopted, which is derived by matching the analytical formula first proposed by \citet{Kewley:2001lr} to SDSS data (see also \citealt{Kewley:2006lr}). These criteria reduce the number of \HII\ regions to 354. For our main analysis, we focus on the 318 \HII\ regions that lie between 0.2 to 1 $\rm R_{25}$ (see Section~4.1). We also correct for dust extinction using the Balmer decrement, assuming H$\alpha$/H$\beta$ = 2.86 and the extinction curve by \citet[][assuming $R_v=3.1$]{Cardelli:1989qy}. To demonstrate the quality of our \HII\ region spectra, we present in Figure~\ref{fig2} sample spectra from four S/N(H$\alpha$) quartiles.

We constrain the oxygen abundances using the \mappingsiv\ photoionization models \citep[][D13]{Dopita:2013qy}. We adopt the Bayesian inference code \izi\ \citep{Blanc:2015qy} to calculate the posterior probability distribution functions given the D13 model grids and extinction-corrected line fluxes. The marginalized mean oxygen abundances reported by \izi\ are used throughout the paper. Typical 1$\sigma$ errors of the abundance from \izi\ are approximately 0.1~dex. Figure~\ref{fig3} shows the abundance map derived from the data cube. In the Appendix, we also present results deriving from different abundance diagnostics and calibrations. The main conclusions of the paper are independent of the oxygen abundance diagnostic.

\begin{figure}[!t]
\centering
\includegraphics[width=\columnwidth]{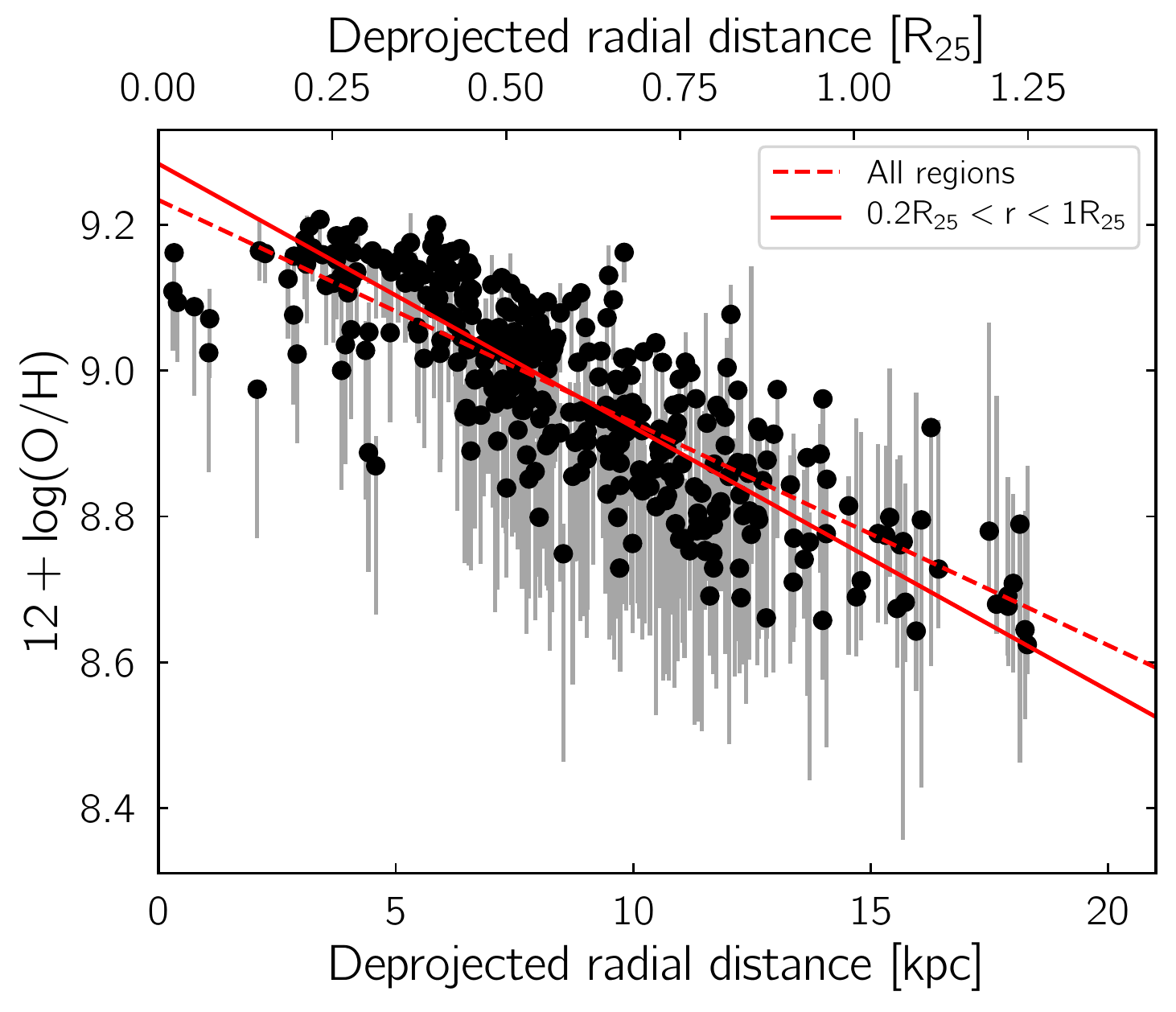}
\caption{Oxygen abundance gradient in NGC\,2997. The black data points correspond to the \HII\ regions shown in Figure~\ref{fig3}. The red lines indicate the linear best-fits to all data points (dashed) and those between 0.2 to 1 $\rm R_{25}$.}\label{fig6} 
\end{figure}

\subsection{Spiral arm mask}
A spiral arm mask is constructed in order to quantitatively compare azimuthal variations of oxygen abundance relative to spiral structures. At least three spiral-like structures can be visually identified in Figure~\ref{fig1}. 
The spiral structures traced by blue, young stars are particularly obvious on the color composite image, and also on the H$\alpha$ image as strings of prominent star-forming regions. Indeed, the previous imaging study by \citet{Vera-Villamizar:2001nr} has found  $m=2$ and $m=3$ modes in NGC\,2997. 

We define spiral arms by visually identifying strings of bright knots on the H$\alpha$ map whose spatial distribution can be approximately described by the logarithmic spiral formula,
\begin{equation}
r(\phi) = r_0 e^{\tan{\theta_p}(\phi-\phi_0)},
\end{equation}
or equivalently,
\begin{equation}
\ln{r(\phi)} = \ln{r_0} + \tan{\theta_p} (\phi-\phi_0).
\end{equation}
Here, $\theta_p$ is the pitch angle. In Figure~\ref{fig4}, we show our H$\alpha$ map in polar coordinates. The deprojected radial distance is shown in logarithmic scale. In this projection, a logarithmic spiral arm would appear to be a straight line. Here, we only focus on deprojected radial distance of $\rm 0.2R_{25}$ to $\rm R_{25}$ (see Section~4.1). By comparing with spiral-like structures seen in Figure~\ref{fig1}, we identify four virtually linear structures in the logarithmic polar projection. The four structures are labeled as S1 to S4 in Figure~\ref{fig4}. Their corresponding locations in the plane of the sky are shown in Figure~\ref{fig5}. The spatial extents of these structures are defined empirically such that the spiral arm mask covers the bulk of the bright H$\alpha$ knots forming the virtually linear structures. In Table~\ref{tbl2}, we tabulate the spiral parameters, including the inner and outer radii and azimuthal angles, pitch angle and azimuthal width.

\begin{table}[!hb]
 \caption{Spiral parameters}
 \label{tbl2}
 \centering
 \begin{tabular}{ccccc}
  \hline
  Structure & ($r_{\rm in},r_{\rm out}$) & ($\phi_{\rm in},\phi_{\rm out}$) & $\theta_p$ & width \\
            & [$\rm R_{25}$] & [$^\circ$] & [$^\circ$] & [$^\circ$]\\
  \hline
  S1 & (0.20, 0.61) & (187.2,  12.6) & 20 & 50 \\
  S2 & (0.20, 0.47) & (320.9, 235.6) & 30 & 35 \\
  S3 & (0.47, 0.90) & (236.9, 199.6) & 45 & 35 \\
  S4 & (0.37, 0.83) & (215.0,  72.2) & 18 & 50 \\
  \hline
 \end{tabular}
\end{table}

\begin{figure}[!t]
\centering
\includegraphics[width=0.95\columnwidth]{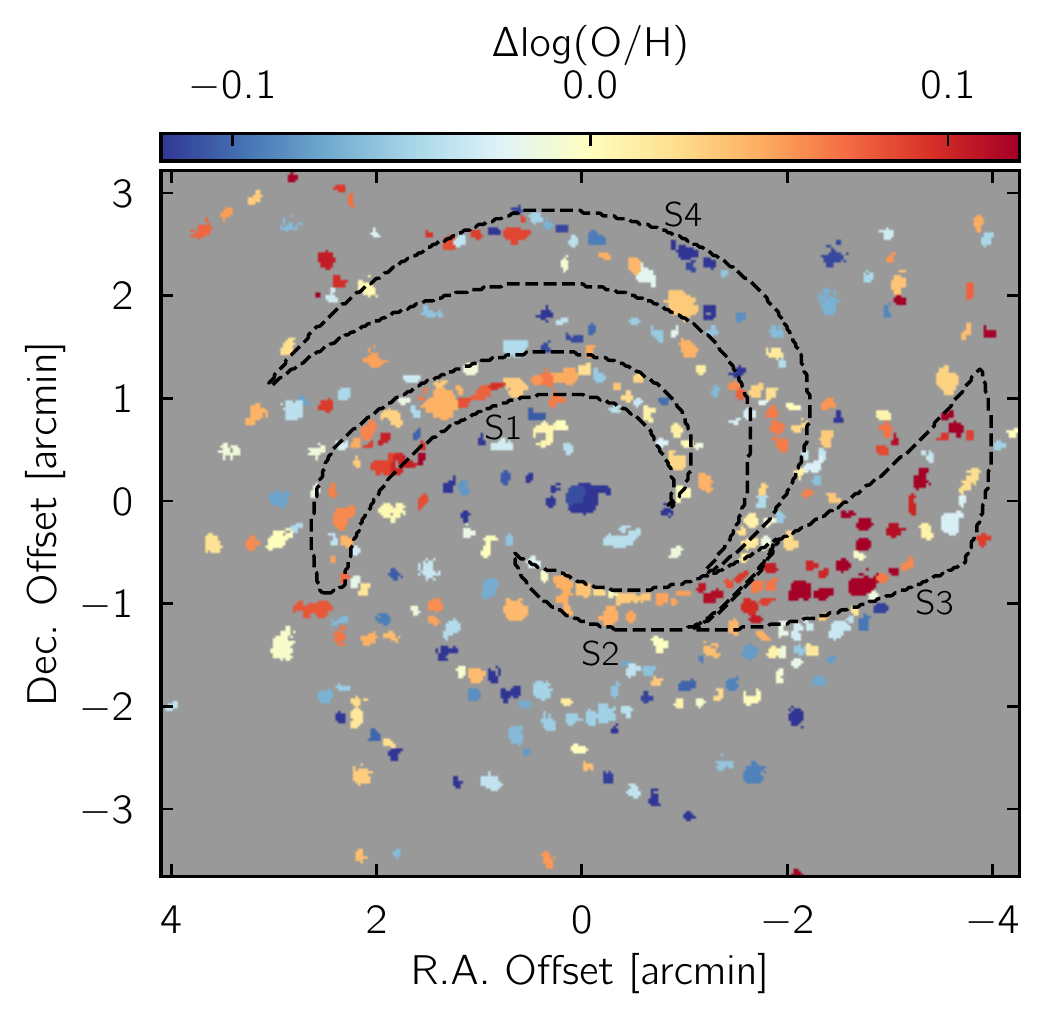}
\caption{Oxygen abundance residual map. The residual map is constructed by removing the best-fit linear gradient (solid line in Figure~\ref{fig6}) from the abundance map (Figure~\ref{fig3}). The spiral structures identified in Figures~\ref{fig4} and \ref{fig5} are also marked as dashed-curves. }\label{fig7} 
\end{figure}

The first spiral structure, S1, corresponds to the main spiral arm in the eastern and northern part of NGC\,2997. S1 can be unambiguously identified due to its brightness and well-behaved spatial distribution that can be characterized by a logarithmic spiral. The pitch angle of S1 is approximately $20^\circ$. The second spiral structure, S2, is offset approximately $180^\circ$ from S1 and has a pitch angle of roughly $30^\circ$. S2 is well defined between $\rm 0.2R_{25}$ and $\rm 0.4R_{25}$. At approximately $\rm 0.4R_{25}$, S2 appears to bifurcate into two separate structures. The S3 structure has a larger pitch angle of $45^\circ$ and continues to extend toward the west. The S4 structure has a smaller pitch angle of $18^\circ$ and spirals outward toward the northern and eastern part of NGC\,2997. 

\begin{figure*}
\centering
\includegraphics[width=0.85\textwidth]{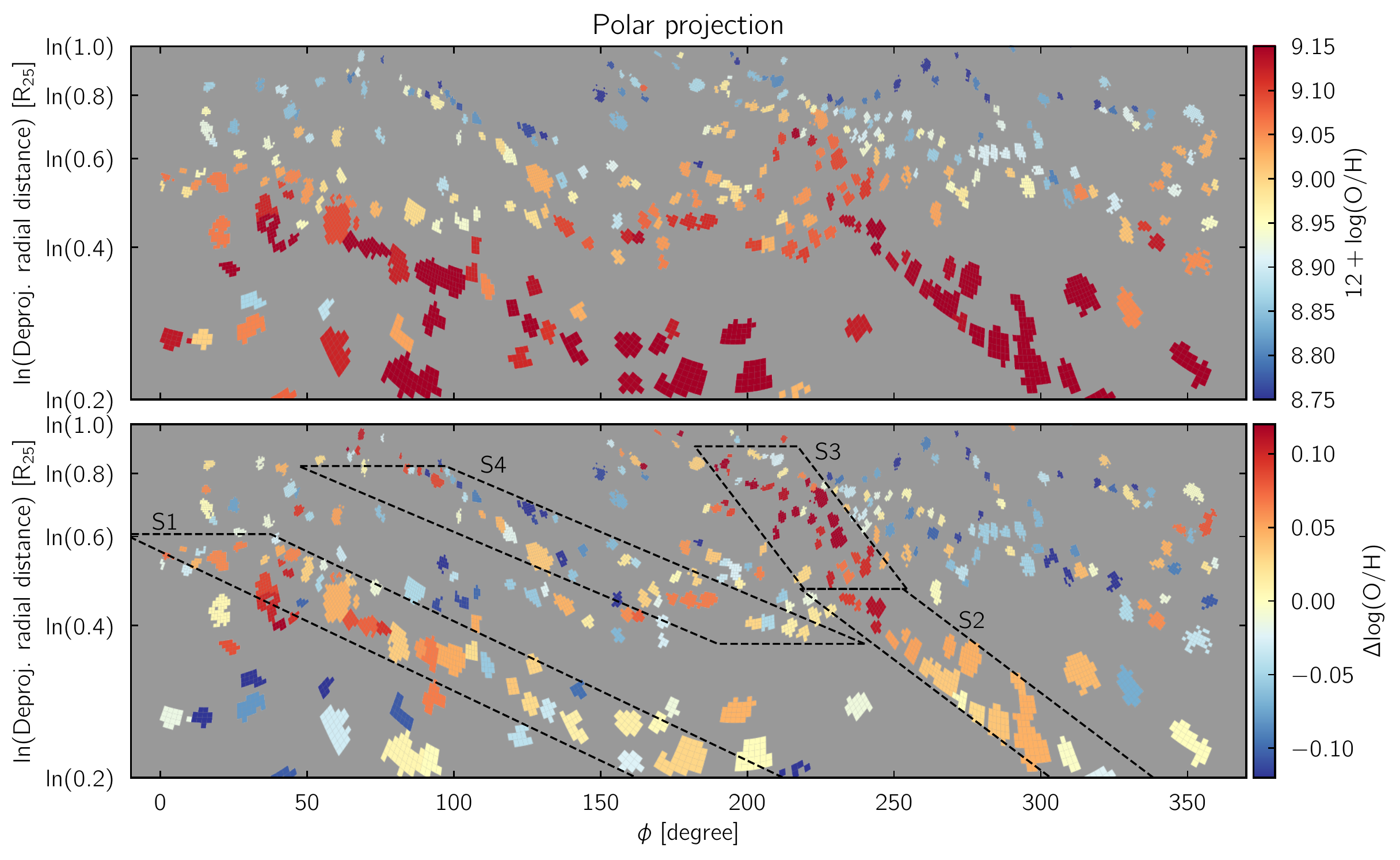}
\caption{Oxygen abundance (upper) and abundance residual (lower) maps, same as Figures~\ref{fig3} and \ref{fig7} but in polar projection. The spiral structures identified in Figures~\ref{fig4} and \ref{fig5} are marked as dashed-curves in the lower panel.}\label{fig8}
\end{figure*}

\begin{table}[!hb]
 \caption{Oxygen abundance gradient}
 \label{tbl3}
 \centering
 \begin{tabular}{cccc}
  \hline
  Slope & Slope & Offset & Scatter \\
  $\rm [dex~kpc^{-1}]$  & $\rm [dex~R_{25}^{-1}]$ & [dex] & [dex]\\
  \hline
  \multicolumn{3}{c}{{All \HII\ regions}} \\
  $(-3.04\pm0.14)\times10^{-2}$ & $-0.45\pm0.02$ & $9.233\pm0.013$ & 0.082\\
\hline
  \multicolumn{3}{c}{{$\rm 0.2R_{25}<r<1R_{25}$}} \\
  $(-3.61\pm0.17)\times10^{-2}$ & $-0.53\pm0.03$ & $9.283\pm0.014$ & 0.080 \\
  \hline
 \end{tabular}
\end{table}

Using the spiral arm mask defined in Figures~\ref{fig4} and \ref{fig5}, we categorise each \HII\ region to either belonging to a spiral arm or in the inter-arm region. Over the deprojected radial distance of $\rm 0.2R_{25}$ to $\rm 1R_{25}$, 41, 18, 33, and 38 regions are assigned to S1, S2, S3, and S4, respectively. The remaining 188 regions are in the inter-arm region. It is worth noting that the four spiral structures contains approximately half of the H$\alpha$ flux. Over the deprojected radial distance of $\rm 0.2R_{25}$ to $\rm 1R_{25}$, about 58\% of the total spaxel H$\alpha$ flux is in the spiral regions (21, 10, 11, and 16\% for S1, S2, S3, and S4, respectively).

\section{Results}
Using the line maps, \HII\ region catalog, and spiral mask described in the previous section, we now investigate the radial gradient and azimuthal variations of oxygen abundance in NGC\,2997.

\subsection{Oxygen abundance gradient}
A prominent radial gradient can be easily identified in the \HII\ region oxygen abundance map (Figure~\ref{fig3}). In Figure~\ref{fig6}, we de-project Figure~\ref{fig3} assuming a circular thin disk and the geometric parameters in Table~\ref{tbl1}. We quantify the abundance radial profile by performing unweighted least-squares linear fits to all the \HII\ regions. The bootstrap data are fit 1,000 times, and the means and standard deviations of the best fits are reported in Table~\ref{tbl3}. The mean best-fit gradient is shown in Figure~\ref{fig6} as the dotted line. 

The best-fit gradient using all \HII\ regions appears to provide a suboptimal description of the data between 0.2 to 1 $\rm R_{25}$. The \HII\ regions at small radii ($r<0.2\rm R_{25}$) fall systematically below the best-fit gradient and at large radii ($r>1\rm R_{25}$) systematically above. Recent observations have demonstrated that this behavior could be quite common in disk galaxies \citep{Sanchez:2014fk,Sanchez-Menguiano:2016vl,Sanchez-Menguiano:2018aa}. To properly quantify the radial gradient over the range that spiral arms and the azimuthal variations are prominent, a second fit is performed using {\em only} \HII\ regions between 0.2 to 1 $\rm R_{25}$. The best-fit is shown in Figure~\ref{fig6} as the solid line. Hereafter, we will only focus on this radial range.

\begin{figure}
\centering
\includegraphics[width=\columnwidth]{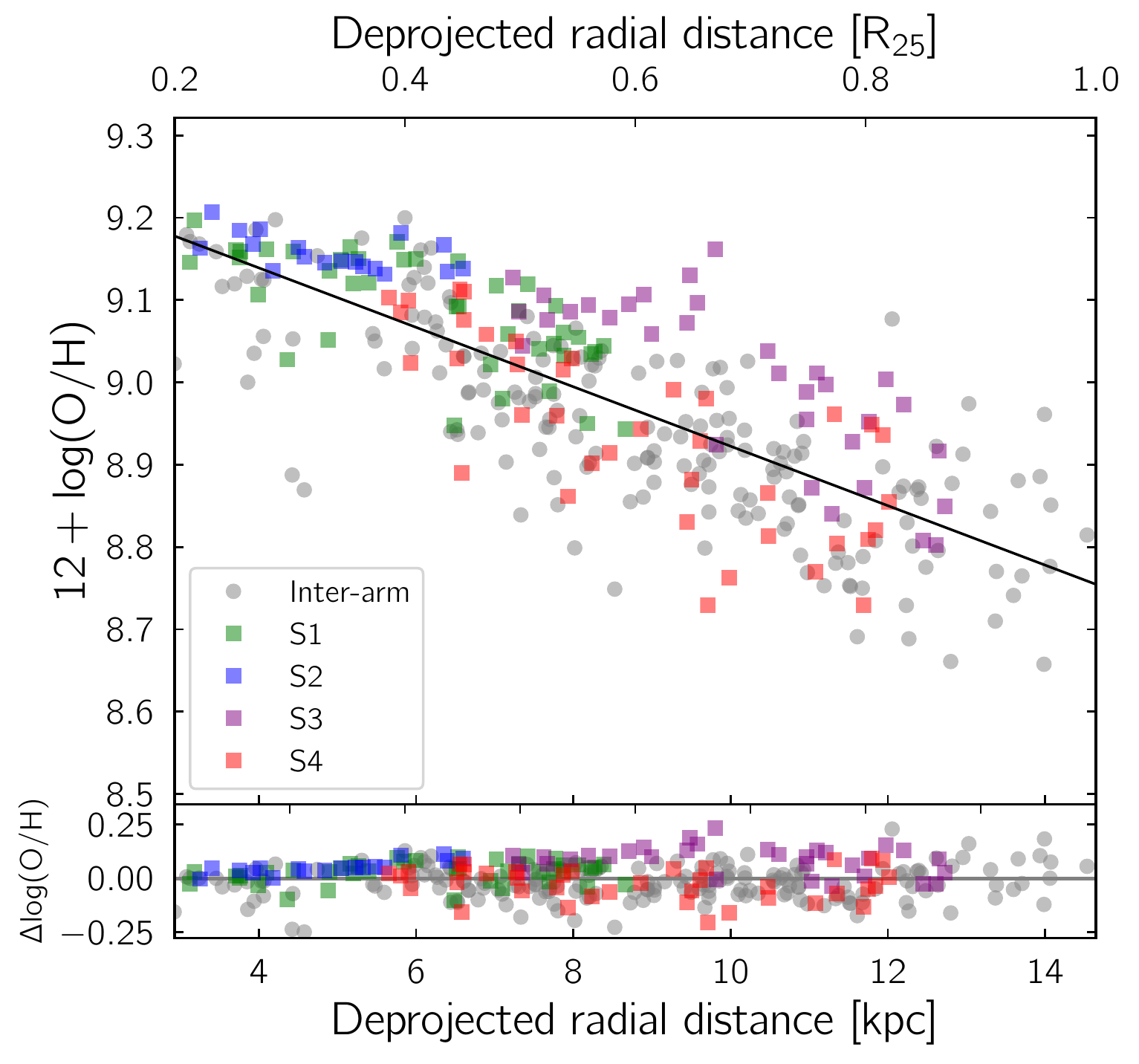}
\caption{Oxygen abundance gradient in NGC\,2997. The different colors correspond to \HII\ regions in different environment. The four spiral arm regions (S1 to S4) are defined in Figures~\ref{fig4} and \ref{fig5}. The solid line is the linear best-fit, same as the one shown in Figure~\ref{fig6}. }\label{fig9}
\end{figure}
\subsection{Azimuthal variations of oxygen abundance and correlation with spiral arms}

\begin{figure*}
\centering
\includegraphics[width=\textwidth]{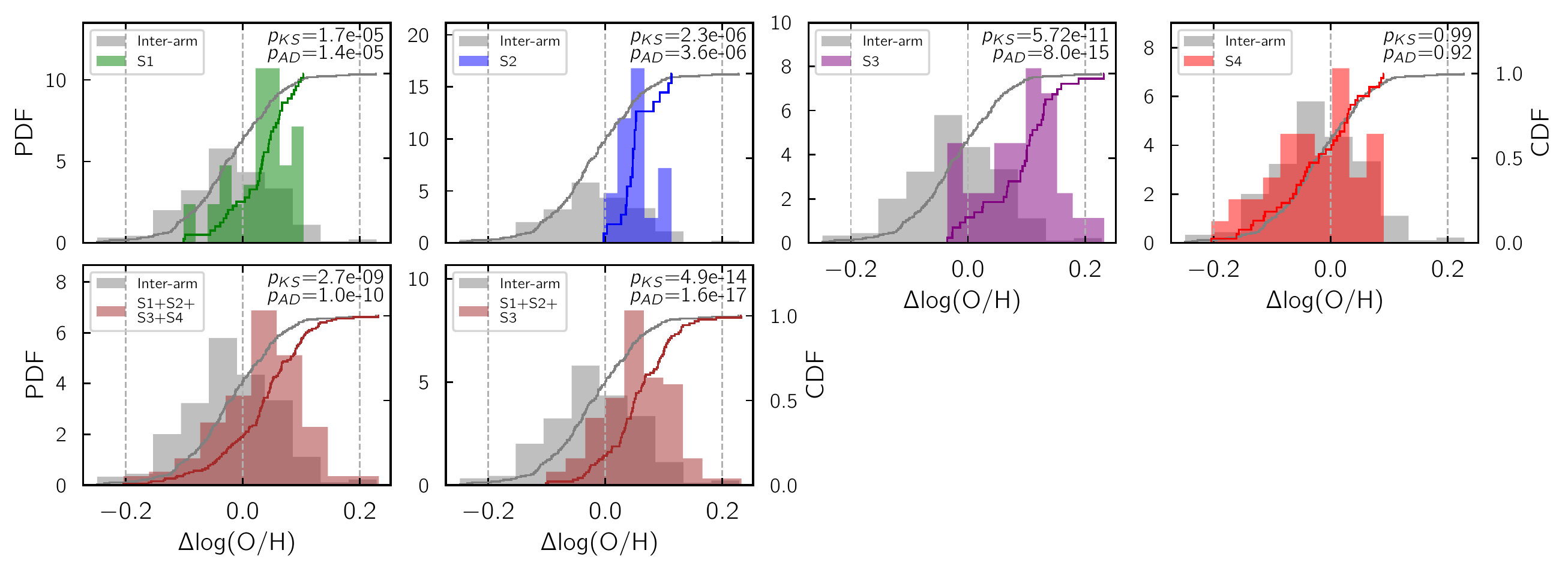}
\caption{ Distributions of abundance residuals ($\rm\Delta\log(O/H)$; best-fit linear gradient subtracted). The histograms correspond to the left-hand-side vertical axes (PDF; probability distribution function),  and the curves correspond to the right-hand-side vertical axes (CDF; cumulative distribution function). The top row compares the inter-arm \HII\ regions with those in the four different spiral arm structures, i.e.~S1 to S4. The bottom row compares the inter-arm \HII\ regions with S1+S2+S3+S4 (all spiral arm structures; first panel from the left) and S1+S2+S3 (without S4; second panel from the left). The {\it p}-values from two-sided Kolmogorov–Smirnov and Anderson-Darling tests are labeled in each panel ($p_{KS}$ and $p_{AD}$). Details are discussed in Section~4.2.}\label{fig10}
\end{figure*}

\begin{figure*}
\centering
\includegraphics[width=\textwidth]{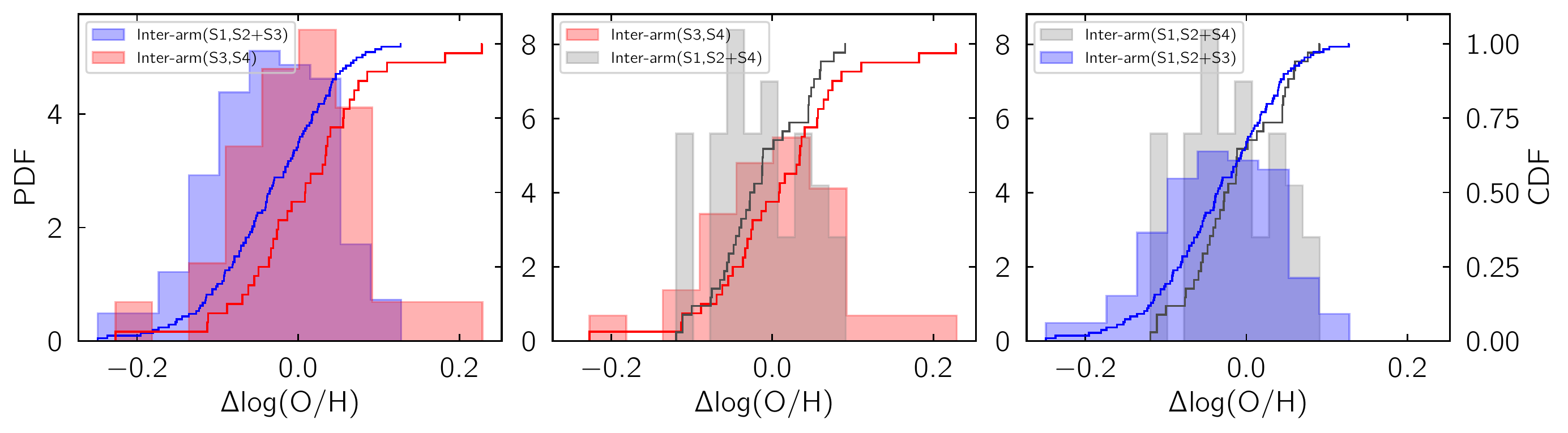}
\caption{Distributions of abundance residuals, $\rm\Delta\log(O/H)$, for different inter-arm regions. The four spiral structures defined in Figures~\ref{fig4} and \ref{fig5} divide the galactic disk into three inter-arm regions. Each of the three panels compares two of the three inter-arm regions. ``Inter-arm(S1,S2+S3)'' denotes the inter-arm region between the S1 and S2+S3 spiral regions. Both histograms (left y-axis) and cumulative distribution functions (right y-axis) are shown. Under the null hypothesis that the two distributions have the same mean, the {\it p-}values derived from bootstrap are tabulated in Table~\ref{tbl4}.}\label{fig11}
\end{figure*}

We remove the best-fit radial gradient from the oxygen abundance map to investigate the azimuthal variations. Here, we adopt the gradient from fitting \HII\ regions between 0.2 to 1 R$_{25}$. The main results of the paper do not change if we adopt a gradient from fitting 0.3 to 1 R$_{25}$ or use boxcar smoothing to derive an average radial profile. The residual map is shown in Figure~\ref{fig7}. In Figure~\ref{fig8}, we also present the abundance and abundance residual maps in the polar projection. The residual maps demonstrate that the scatter around the best-fit abundance gradient is not random. Rather, a pattern that seemingly correlates with the spiral structures can be recognised. In Figures~\ref{fig7} and \ref{fig8}, we also mark the four spiral structures identified in Section~3.2. The residuals appear to be systematically positive in the spiral arm regions, particularly for S1, S2, and S3.

We quantitatively compare abundance residuals with the spiral structures by investigating the distributions of the residuals. In Figure~\ref{fig9}, we present the abundance gradient and abundance residuals in 1D, and colour-code the \HII\ region based on their environmental categories. The distributions of the abundance residuals, $\rm \Delta\log(O/H)$, are analysed in Figure~\ref{fig10}. In the top row, we compare the distributions of the residuals between the four spiral structures and the inter-arm region. We present both histograms and the cumulative distribution functions (CDFs). Two-sided Kolmogorov-Smirnov (KS) and Anderson-Darling (AD) tests\footnote{In hypothesis testing, a {\it p}-value $<0.05$ is usually considered statistically significant and a {\it p}-value $<0.01$ highly significant.} are also conducted to quantitatively assess whether two distributions are drawn from the same parent distribution.

Figure~\ref{fig10} suggests that, for S1, S2, and S3, the residuals are {\it different} between the inter-arm regions and the spiral arm regions. The low {\it p}-values ($\ll 0.01$) effectively rule out the null hypothesis that the two distributions are drawn from the same parent distribution. The differences between the median $\rm \Delta\log(O/H)$ are 0.054, 0.067, 0.120~dex for S1, S2, and S3, respectively. 

For S4, the statistical tests cannot rule out the null hypothesis that the distributions are the same, which is in sharp contrast to the other three spiral structures. The different chemical property of S4 may imply that S2+S3 is the counterpart of S1 in a 2-arm spiral system, and S4 is in fact a separate structure. 

In the bottom row of Figure~\ref{fig10}, we repeat the comparison but now compare all spiral arm \HII\ regions combined (S1+S2+S3+S4) with the inter-arm \HII\ regions. As expected, the K-S test rules out the null hypothesis. The difference between the medians is 0.056~dex. If we remove S4 from the comparison (assuming that S4 is not a part of the 2-arm system composed of S1, S2, and S3), we reach the same conclusion that the residuals have different distributions (the difference between the medians is 0.070~dex). 

We note that although we only use the D13 abundance calibration in the main text, we repeat the analysis using three different calibrations in the Appendix and reach qualitatively the same conclusion.

\subsection{Azimuthal variations of oxygen abundance between inter-arm regions}

In addition to the variations between arm and inter-arm regions, we now investigate the variations between {\em different} inter-arm regions. The four spiral structures divide the galactic disk into three separate inter-arm regions, i.e. those between S1 and S2+S3, between S3 and S4, and between S1 and S2+S4. We categorise the inter-arm \HII\ regions based on their locations. Ambiguous regions are ignored in the following analysis (e.g.~those lie along the spiral trajectories but outside the defined spiral structure boundaries). The sample sizes are 109, 32, and 34 for Inter-arm(S1,S2+S4), Inter-arm(S3,S4), and Inter-arm(S1,S2+S3), respectively.

In Figure~\ref{fig11}, we present the distributions of the residuals for the three different groups of inter-arm \HII\ regions. To evaluate whether the different groups have statistically different mean $\rm\Delta\log(O/H)$, we perform 2-sample hypothesis tests using the bootstrap technique \citep{book:1130259}. We compare 100,000 bootstrap samples and compute 1-sided {\it p}-values for the null hypothesis that the means of the two groups are the same. The {\it p}-values are 0.007, 0.100, and 0.060, respectively, for comparing between Inter-arm(S1,S2+S4) and Inter-arm(S3,S4), Inter-arm(S3,S4) and Inter-arm(S1,S2+S3), Inter-arm(S1,S2+S4) and Inter-arm(S1,S2+S3), respectively (see Table~\ref{tbl4}). The statistical tests suggest that \HII\ regions in Inter-arm(S1,S2+S4) have on average lower abundances than those in Inter-arm(S3,S4). The test statistics for Inter-arm(S1,S2+S4) being less chemically enriched than Inter-arm(S1,S2+S3) is close to the borderline of significance. 

This comparison, however, depends on the abundance calibration adopted. In Table~\ref{tbl4} and the appendix, we present the same analysis using three different calibrations from \citet[][P04]{Pettini:2004lr}, \citet[][M13]{Marino:2013vn} and \citet[][D16]{Dopita:2016fk}. The P04 and M13 calibrations (which are similar and based on the same diagnostic lines) infer no statistical significant differences between all three inter-arm regions. The D16 calibration, however, gives a similar conclusion as the D13 calibration adopted above. The D16 calibration gives even higher significance values for not only between Inter-arm(S1,S2+S4) and Inter-arm(S3,S4), but also between Inter-arm(S1,S2+S4) and Inter-arm(S1,S2+S3). 

\begin{table}[]
\caption{{\it p-}value of 2-sample hypothesis test}
 \label{tbl4}
 \centering
 \begin{tabular}{llllll}
  \hline
 \multicolumn{2}{c}{Inter-arm regions}  & D13$^a$ & P04$^b$ & M13$^b$ & D16$^b$ \\
  \hline
(S1,S2+S4) & (S3,S4) & {0.007$^c$} & {0.211} & {0.214} & {0.004$^c$} \\
(S3,S4) & (S1,S2+S3) & {0.100} & {0.353} & {0.353} & {0.132} \\
(S1,S2+S4) & (S1,S2+S3) & {0.060} & {0.308} & {0.309} & {0.011$^c$} \\
\hline
\multicolumn{6}{p{1\columnwidth}}{
\footnotesize The null hypothesis is that the two distributions have the same mean. }\\
\multicolumn{6}{p{1\columnwidth}}{
\footnotesize 
$^a$~Calibration used in the main text. 
$^b$~Calibrations presented in the appendix. 
$^c$~Considered as statistically significant ({\it p-}value$<0.05$).}

\end{tabular}
\end{table}

To summarise, our data support the presence of azimuthally varying oxygen abundances. At the same galactocentric distance, \HII\ regions in spiral arm regions are likely to have higher abundances (0.06 to 0.07~dex) than those in the inter-arm region. Between different inter-arm regions, Inter-arm(S1,S2+S4) appears to have the lowest abundances, although this conclusion may change depending on the adopted calibrations.

\section{Discussion}

\begin{figure*}
\centering
\includegraphics[width=\textwidth]{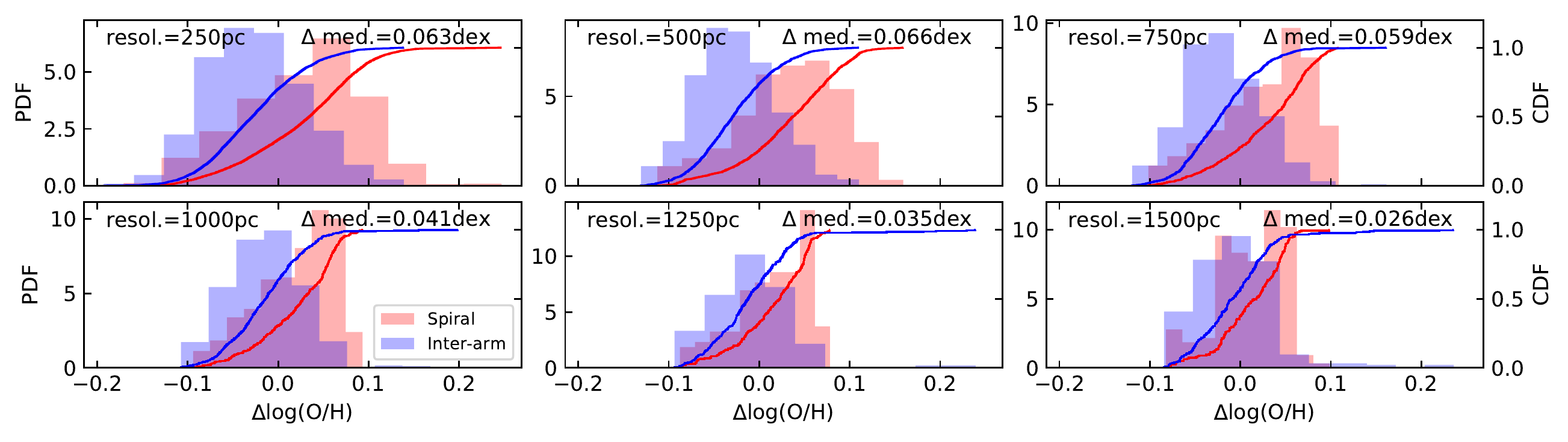}
\caption{Distributions of the abundance residuals, $\rm\Delta\log(O/H)$, for the spiral and inter-arm spaxels (red and blue, respectively). Different panels correspond to convolving the TYPHOON cube to different spatial resolutions, as labeled in the upper left corner of each panel. Both histograms and the cumulative distribution functions are shown. The difference between the medians is labeled in the upper right corner of each panel. }\label{fig12}
\end{figure*}

\begin{figure*}
\centering
\includegraphics[width=0.8\textwidth]{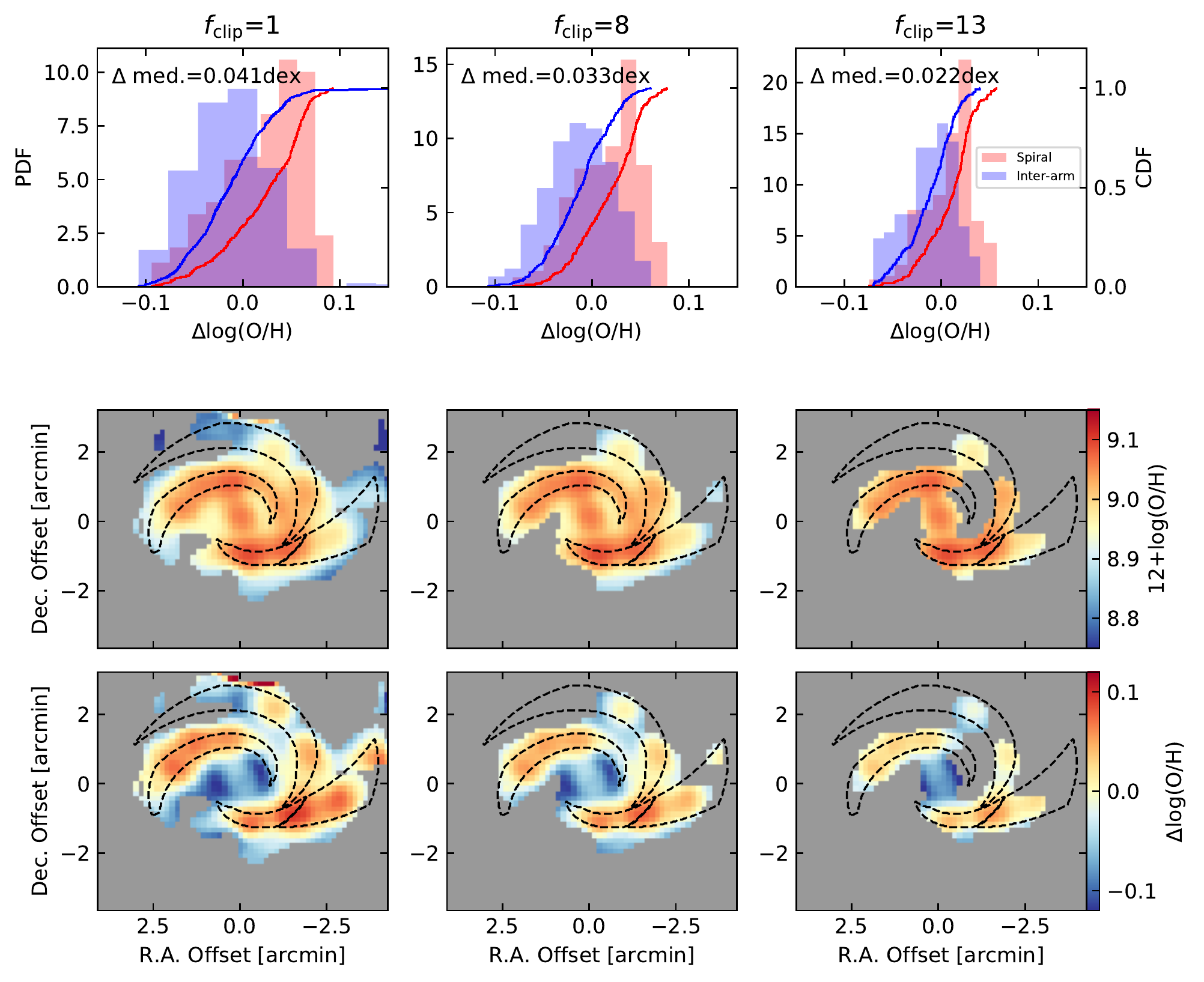}
\caption{Simulation of the effect of signal-to-noise ratio on the observed azimuthal variations. 
Each column corresponds to one set of simulation using different $f_{\rm clip}$ values. A higher $f_{\rm clip}$ corresponds to applying a higher H$\alpha$ threshold on spaxels entering the analysis, i.e.~poorer signal-to-noise data. The starting point of the analysis (left column) is at 1000~pc resolution, same as the lower bottom panel of Figure~\ref{fig12}. The different rows present the abundance residual distributions (top), the abundance maps (middle), and the abundance residual maps (bottom). The observed azimuthal variations decreases with increasing  $f_{\rm clip}$ (decreasing signal-to-noise), }\label{fig13}
\end{figure*}

Recent studies have found azimuthal variations of \HII\ region oxygen abundance imprinted on top of radial gradients that are commonly seen in local star-forming disks \citep{Sanchez-Menguiano:2016zr,Vogt:2017aa,Ho:2017aa,Sakhibov:2018aa}. These discoveries are driven by the increasing numbers of 3D spectroscopy observations of nearby galaxies. Unlike traditional long-slit observations that target only selected \HII\ regions, these new observations deliver a complete view of the oxygen abundance distributions, facilitating the discovery of azimuthal variations. The increasing number of such discoveries naturally gives rise to the question of how common azimuthal variations are in the nearby Universe. 

To address the prevalence of the azimuthal variations in star-forming disks, a critical first step was taken by \citet{Sanchez-Menguiano:2017aa}. They analyzed a sample of 63 face-on spiral galaxies from the CALIFA survey. They visually traced spiral arms on SDSS {\em g-}band images and fit abundance gradients for spaxels belonging to the spiral arm or inter-arm regions. The sample was divided into two groups, flocculent and grand-design spirals, by visual inspection, basing on the scheme proposed by \citet{Elmegreen:1982ab} and \citet{Elmegreen:1987aa}. \citeauthor{Sanchez-Menguiano:2017aa} found subtle but statistically significant differences between the arm and inter-arm abundance gradients for their flocculent galaxies, but not for the grand-design spirals.

So far, the TYPHOON Program has reported two nearby galaxies (within 20~Mpc) that exhibit pronounced azimuthal variations: NGC\,2997 reported in this work and NGC\,1365 by \citet{Ho:2017aa}. Surprisingly, these two galaxies both present grand-design spirals. Indeed, the two galaxies are categorized by \citet{Elmegreen:1987aa} as AC9 and AC12, respectively. These categories are considered as ``grand-design'' in the binary classification of \citet{Sanchez-Menguiano:2017aa}. The grand-design morphologies of NGC\,1365 and NGC\,2997 appear to contradict the finding of \citeauthor{Sanchez-Menguiano:2017aa} that arm to inter-arm variations are more common in flocculent galaxies. However, the sample size of two is too low to be considered as statistically significant. 

The CALIFA survey and the TYPHOON Program reach distinctly different spatial resolutions. The CALIFA survey has a typical spatial resolution of 1~kpc, significantly larger than the typical \HII\ region sizes. At the much higher spatial resolution of 100 to 200~pc reached by the TYPHOON Program, \HII\ regions can be marginally resolved and isolated into individual bins, allowing more accurate estimates of the oxygen abundances. This difference leads to the question of to what extent the spatial resolution impacts the detectability of the azimuthal variations. 

To approach this question, we convolve the original TYPHOON data cube to resolutions of 250, 500, 750, 1000, 1250, and 1500~pc (FWHM). The data cubes are re-sampled to grid sizes that equal to half of the resolution FWHM. We then analyse the data cubes following the same procedures described in Sections~3 and 4. The only difference is that we perform spaxel-to-spaxel analysis, same as \citet{Sanchez-Menguiano:2017aa}, instead of identifying and binning \HII\ regions. Only pixels with line ratios consistent with photoionisation by OB stars are considered. Similarly, we remove pixels with S/N less than three in any of the diagnostic lines (see Section~3.1). We adopt the same environmental mask constructed in Section~3.2 using the TYPHOON cube at its native resolution.

In Figure~\ref{fig12}, we compare $\rm\Delta\log(O/H)$ of arm and inter-arm regions using data of different spatial resolutions. Figure~\ref{fig12} shows that the magnitudes of the arm to inter-arm variations decrease with increasing spatial resolution. At the resolutions of 1000 and 1500~pc, the logarithmic differences are approximately 30\% and 50\% lower than that at the naive resolution (110 to 140~pc), respectively. 

The analysis above only considers the effect of resolution. No additional noise was injected into the data cubes. However, the signal-to-noise of the data could further impact our ability to recover the underlying azimuthal variations. We simulate such impact using the 1000~pc data cube. First, we identify the spaxel that enters the previous analysis and has the smallest H$\alpha$ flux value (surface brightness of $3.5\times 10^{-17}\rm~ergs~s^{-1}~cm^{-2}~arcsec^{-2}$). By construction, the spaxel must have line ratios consistent with star formation and S/N of at least three on all the diagnostic lines. Using this spaxel as a baseline, we then apply different H$\alpha$ clippings based on scaling this minimum H$\alpha$ flux value. The scaling factor is denoted as $f_{\rm clip}$. 
After clipping spaxels below the scaled H$\alpha$ thresholds, the same analysis described in Sections~3 and 4 are repeated on a spaxel-to-spaxel basis. 

The resulting comparison between the spiral and inter-arm regions is shown in the top row of Figure~\ref{fig13} for $f_{\rm clip}=$ 1, 8, and 13. We also show the corresponding abundance and abundance residual maps in the middle and bottom rows. Figure~\ref{fig13} suggests that the magnitude of the variations also depends on the signal-to-noise of the data. The variations decrease with decreasing signal-to-noise. In particular, the inter-arm regions suffer from low signal-to-noise more than the arm regions. At $f_{\rm clip}=13$, a significant fraction of inter-arm spaxels is below the noise level. For reference, the numbers of (arm, inter-arm) pixels are (378, 566), (291, 340), and (212, 153) for $f_{\rm clip}$ of 1, 8, and 13, respectively. Note that the $f_{\rm clip}$ values adopted here are chosen to demonstrate the trend of how decreasing data quality impacts the measured azimuthal variations. It is not the goal of these simulations to match the data qualities of existing IFS surveys, which would require a more careful control of the noise characteristics. 

The biases of measuring the azimuthal variations in Figures~\ref{fig12} and \ref{fig13} are the end product of two artefacts: 1) at poor spatial resolutions the diffuse ionised gas contaminating the \HII\ region flux measurements and 2) at low signal-to-noise the drop-off of low brightness \HII\ regions. The second artefact can be further enhanced by a larger beam since the flux from an \HII\ region is redistributed to a larger area. \HII\ regions with lower luminosities are thus more susceptible to both effects, as can be seen in the inter-arm regions in Figure~\ref{fig13}. These artefacts are also known to introduce systematic errors in metallicity gradient measurements \citep{Yuan:2013gf,Mast:2014lr}.

Finally, the analysis above is idealised and does not take into account the difficulties in defining spiral arms using low resolution data. The spiral arm mask derived from the original TYPHOON data was adopted throughout the analysis (Figures~\ref{fig4} and \ref{fig5}). However, tracking the spiral arms with lower resolution images could be non-trivial. Errors in tracking the arms are likely to further dilute the azimuthal variation signal, although in some cases existing data at other wavelengths (e.g. \HI\ or CO) could facilitate the spiral tracking. It is beyond the scope of this work to explore this part of the parameter space. Nevertheless, it is evident that censusing nearby galaxies at a scale approaching the typical scale of \HII\ regions is crucial for quantifying both the prevalence and magnitude of the azimuthal variations of oxygen abundance in star-forming disks

\section{Summary}

We have analysed the \HII\ region oxygen abundances in NGC\,2997 using 3D spectroscopy data from the TYPHOON Project. Working at a spatial resolution of about 125~pc, we have identified about 350 \HII\ regions and complexes, and derived their oxygen abundances using strong emission lines and strong-line diagnostics. 

NGC\,2997 presents a radial abundance gradient similar to those commonly seen in star-forming disks. After the besti-fit linear gradient is subtracted from the abundance map, we find pronounced, 0.06~dex azimuthal variations of oxygen abundance. At a given radius, the \HII\ region oxygen abundances are typically higher on the spiral arms and lower in the inter-arm regions. Three out of the four identified spiral features show such characteristics. The variations are qualitatively the same for all the four abundance calibrations adopted in this work. We also compared the oxygen abundance residuals (i.e.~linear gradient subtracted) between the three different inter-arm regions that are partitioned by the spiral arms. 
Although there is weak evidence that the residuals are not the same between different inter-arm regions, we are unable to confirm this finding with all the adopted calibrations.

We also simulated the detectability of the azimuthal variations if NGC\,2997 were observed at lower spatial resolution and at lower signal-to-noise ratios. We showed that the detectability and magnitude of the variations depends on both the spatial resolution and precision of the data. In general, the underlying azimuthal variations become increasingly difficult to recover from low resolution and shallow data. Deep optical IFU data with spatial resolutions reaching the typical scale of \HII\ regions are required to assess the prevalence and magnitude of azimuthal variations in spiral galaxies. 

\begin{acknowledgements}
This paper includes data obtained with the du Pont Telescope at the Las Campanas Observatory, Chile as part of the TYPHOON Program, which has been obtaining optical data cubes for the largest angular-sized galaxies in the southern hemisphere. We thank past and present Directors of The Observatories and the numerous time assignment committees for their generous and unfailing support of this long-term program. 
We also thank the referee for her/his constructive comments that improve the quality of this work. ITH is grateful for the support of an MPIA Fellowship. SEM acknowledges funding from the Deutsche Forschungsgemeinschaft (DFG) vsia grant SCHI 536/7-2 as part of the priority program SPP 1573 "ISM-SPP: Physics of the Interstellar Medium.

\end{acknowledgements}

\begin{appendix}

\section{Oxygen abundance derived from other calibrations}
In the main part of the paper, we derive the oxygen abundance using the D13 photoionization models. Here, we use three different diagnostics and calibrations to derive oxygen abundances. Key analyses of the paper are then repeated. 

We calculate oxygen abundances using the O3N2 index: 
\begin{equation}
{\rm O3N2} = \log \Big({[\textrm{O}\textsc{iii}]\lambda 5007 / {\rm H\beta} \over [\textrm{N}\textsc{ii}]\lambda 6583/ {\rm H\alpha}}\Big). 
\end{equation}
Two calibrations by \citet[][P04]{Pettini:2004lr} and \citet[][M13]{Marino:2013vn} are adopted. We also calculate oxygen abundances using the N2S2 calibration by \citet[][D16]{Dopita:2016fk} based on the [\ion{N}{2}]/H$\alpha$ and [\ion{S}{2}]/H$\alpha$ line ratios. The N2S2 calibration is based on the \mappingsv\ code, an updated to \mappingsiv\ used in D13.

\begin{figure*}
\centering
\includegraphics[width=0.9\textwidth]{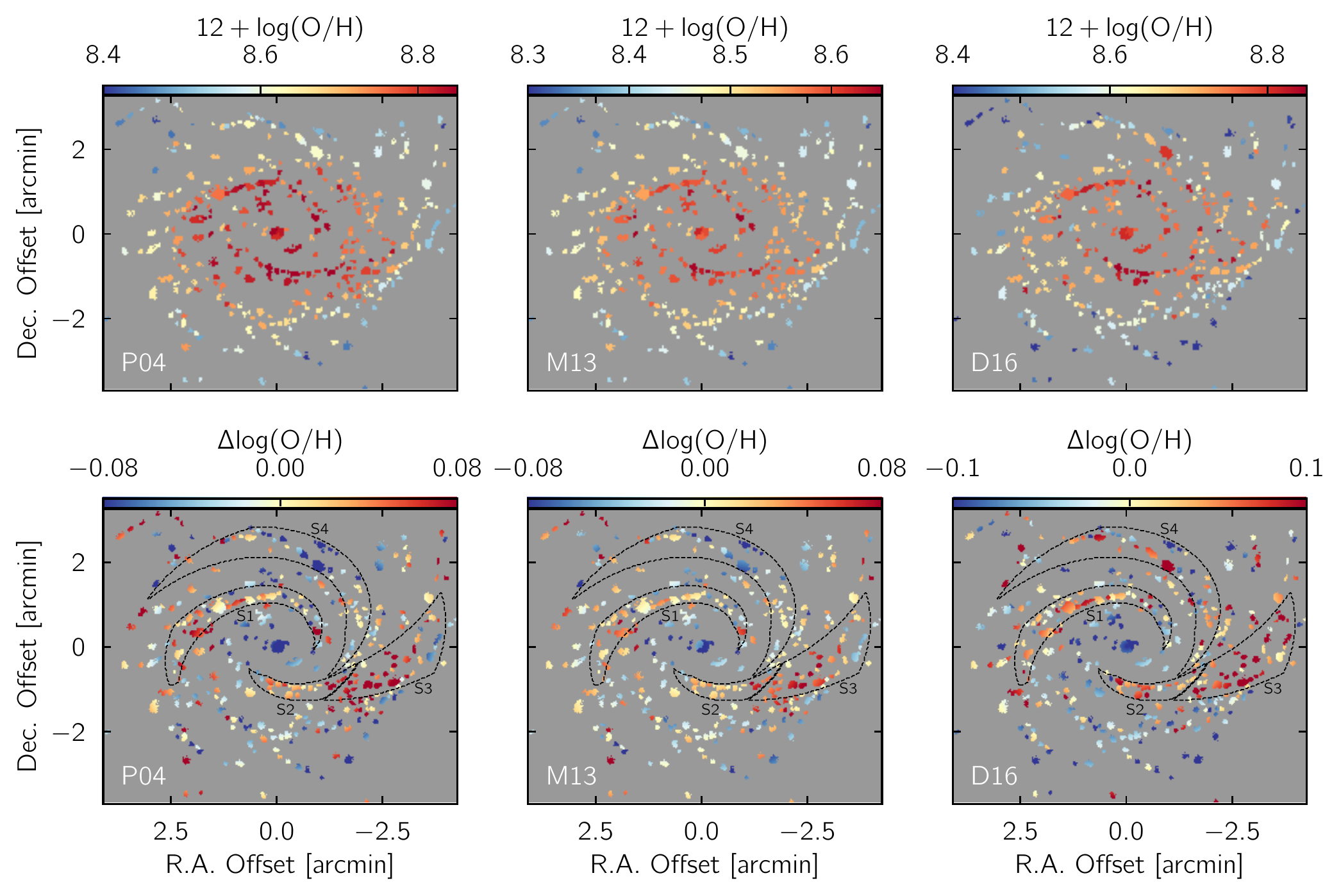}
\caption{Similar to Figures~\ref{fig3} and \ref{fig7}, but for different abundance calibrations. Upper panels: oxygen abundance maps. Lower panels: oxygen abundance residuals.}\label{figa1}
\end{figure*}

\begin{figure*}
\centering
\includegraphics[width=1\textwidth]{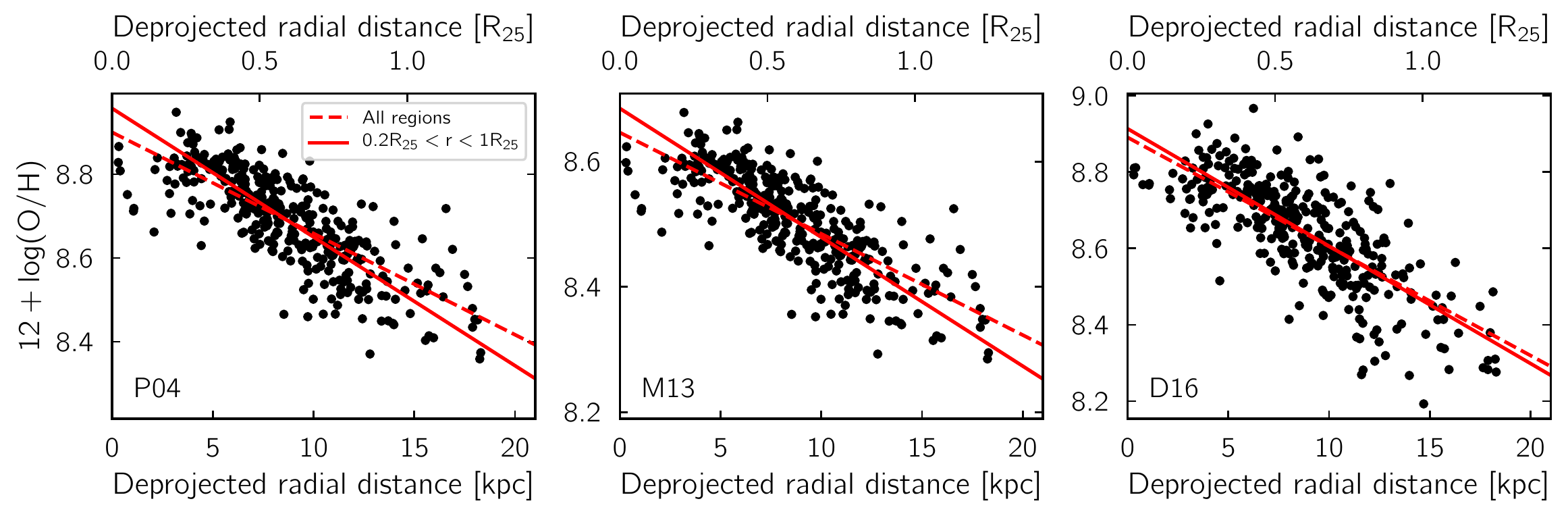}
\caption{Similar to Figure~\ref{fig6} but for different abundance calibrations. The best-fit gradients are tabulated in Table~\ref{tbla1}.}\label{figa2}
\end{figure*}

In Figure~\ref{figa1}, we present the oxygen abundance maps (top rows) and oxygen abundance residual maps (bottom rows) using the three different calibrations. We present the oxygen abundance gradients in Figure~\ref{figa2} and tabulate the best-fit abundance gradients in Table~\ref{tbla1}. 

\begin{figure*}
\centering
\includegraphics[width=0.9\textwidth]{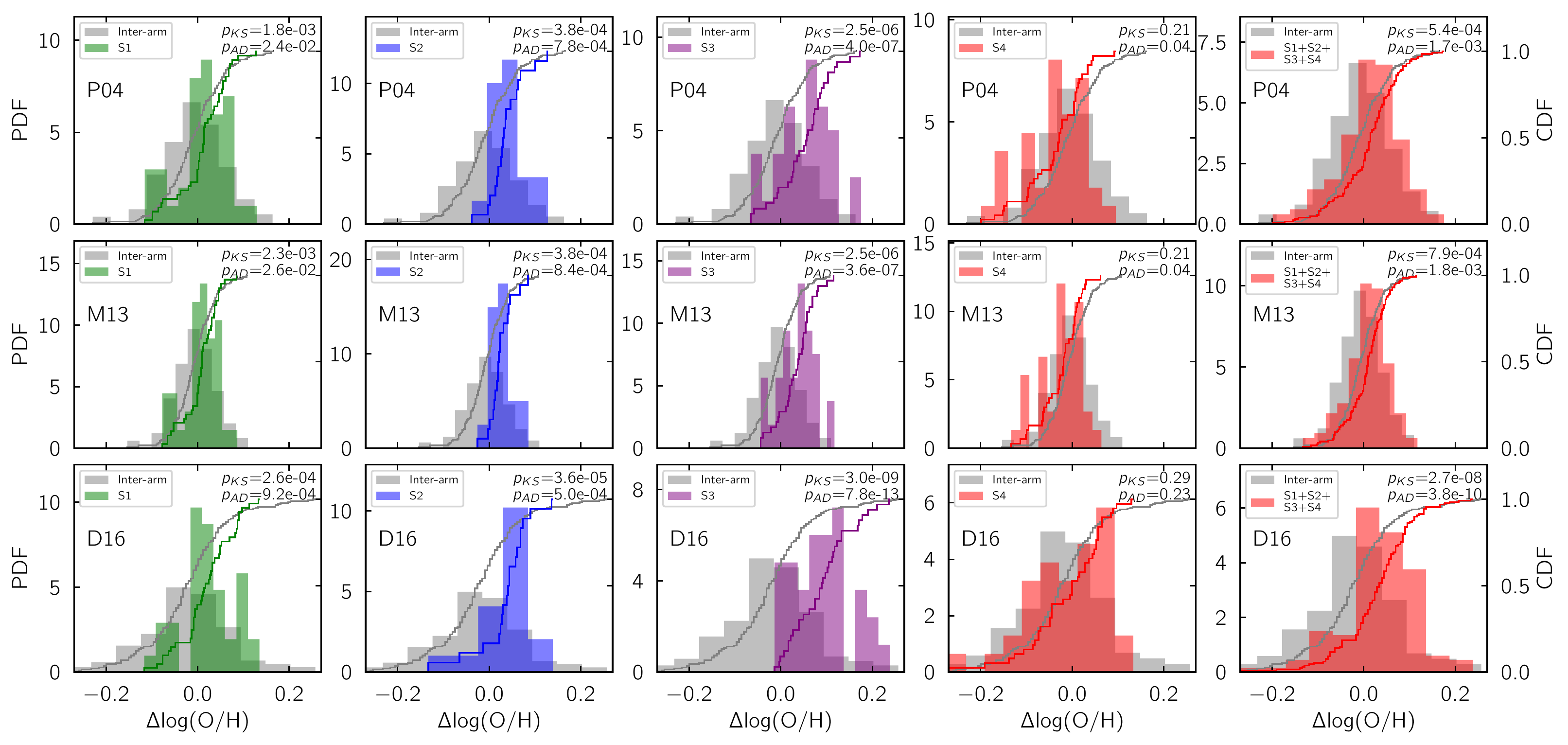}
\caption{Same as Figure~\ref{fig10}, but for different abundance calibrations.}\label{figa3}
\end{figure*}

In Figure~\ref{figa3}, we repeat the analysis of Figure~\ref{fig10}. We compare abundance residuals between the four spiral structures and the inter-arm region. We reach the same conclusion that S1, S2, and S3 have abundance residuals significantly different from the inter-arm region. 
This is supported by both the KS and AD tests. For S4, the KS test suggests that S4 and the inter-arm region have similar distributions. The {\it p}-values of the AD test, however, are close to the border line of significance for the P04 and M13 calibrations. 

When we combine all spiral arm regions and compare them with the inter-arm regions (right column of Figure~\ref{figa3}), we reach the same conclusion that the spiral arms have on average higher abundances than the inter-arm regions. The differences between the medians are 0.026, 0.017, and 0.058~dex, for P04, M13, and D16, respectively. This is to be compared with 0.056~dex derived from D13 and \izi\ in the main text. This discrepancy is unsurprising given that the abundance differences between two calibrations are often more complicated than a simple offset. The differences in abundance are a function of abundance \citep{Kewley:2008qy}. Hence, it remains non-trivial to quantify the degree of the azimuthal variations to high accuracy.

\begin{figure*}
\centering
\includegraphics[width=0.9\textwidth]{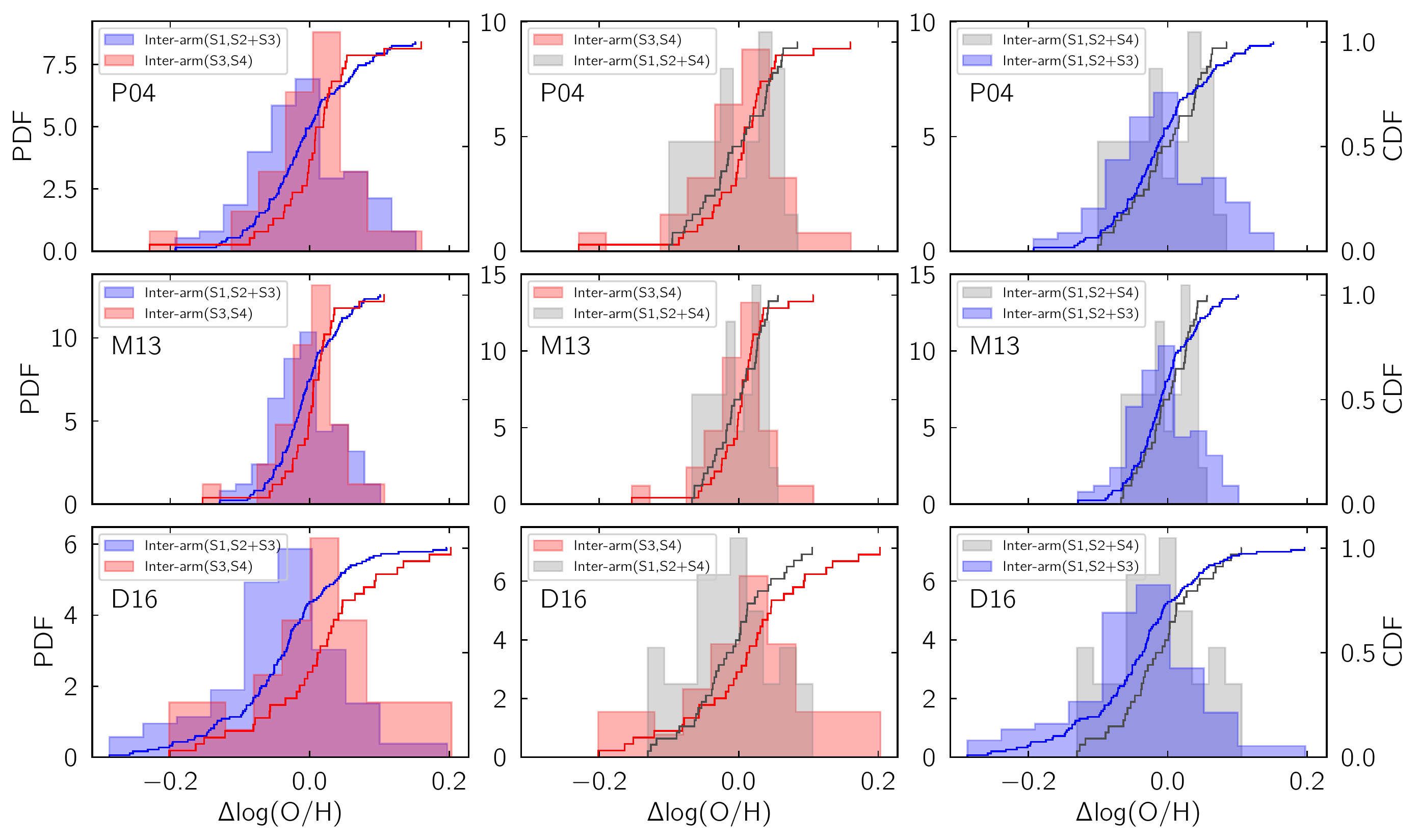}
\caption{Same as Figure~\ref{fig11}, but for different abundance calibrations. The {\it p-}values for the hypothesis test are listed in Table~\ref{tbl4}. The null hypothesis is that the two distributions have the same mean value.}\label{figa4}
\end{figure*}

In Figure~\ref{figa4}, we repeat the analysis of Figure~\ref{figa4}, comparing abundance residuals between different inter-arm regions. The same hypothesis test is performed using the bootstrap technique, under the null hypothesis that the means of the two distributions are the same. The resulting {\it p-}values are tabulated in Table~\ref{tbl4}. Here, the P04 and M13 calibrations yield no statistically significant differences between all inter-arm regions. The D16 calibration yields a similar result to the D13 calibration adopted in the main text. The difference between Inter-arm(S1,S2+S4) and Inter-arm(S1,S2+S3) is only marginally significant for the D13 calibration ({\it p-}value of 0.060) but significant for the D16 calibration ({\it p-}value of 0.011). 

\begin{table*}[!hb]
  \caption{Oxygen abundance gradients for different calibrations}
 \label{tbla1}
\centering
 \begin{tabular}{ccccc}
  \hline
  & Slope & Slope  & Offset & Scatter \\
 & [$\rm dex~kpc^{-1}$]  & [$\rm dex~R_{25}^{-1}$] & [dex] & [dex]\\
  \hline
  \multicolumn{4}{c}{{All \HII\ regions}} \\  
P04  & $(-2.42\pm0.17)\times10^{-2}$ & $-0.35\pm0.02$ & $8.900\pm0.014$ & 0.074\\
M13  & $(-1.61\pm0.11)\times10^{-2}$ & $-0.24\pm0.02$ & $8.647\pm0.010$ & 0.049\\
D16  & $(-2.86\pm0.16)\times10^{-2}$ & $-0.42\pm0.02$ & $8.891\pm0.013$ & 0.091\\
\hline
  \multicolumn{4}{c}{{$\rm 0.2R_{25}<r<1R_{25}$}} \\
P04  & $(-3.08\pm0.13)\times10^{-2}$ & $-0.45\pm0.02$ & $8.958\pm0.011$ & 0.066\\
M13  & $(-2.06\pm0.09)\times10^{-2}$ & $-0.30\pm0.01$ & $8.685\pm0.008$ & 0.044\\
D16  & $(-3.07\pm0.20)\times10^{-2}$ & $-0.45\pm0.03$ & $8.914\pm0.016$ & 0.089\\
\hline
 \end{tabular}
\end{table*}

\end{appendix}

\bibliographystyle{aa} 
\bibliography{references} 
\end{CJK*}

\end{document}